# Two-dimensional superconductivity at the interface of a $Bi_2Te_3$/FeTe heterostructure

Qing Lin He[2,3]†, Hongchao Liu[1]†, Mingquan He[1,2], Ying Hoi Lai[2,3], Hongtao He[1,4], Gan Wang[4], Kam Tuen Law[1], Rolf Lortz[1,2], Jiannong Wang[1,2] and Iam Keong Sou[1,2,3]*

[1] Department of Physics, the Hong Kong University of Science and Technology, Clear Water Bay, Hong Kong SAR, China

[2] William Mong Institute of Nano Science and Technology, the Hong Kong University of Science and Technology, Clear Water Bay, Hong Kong SAR, China

[3] Nano Science and Technology Program, the Hong Kong University of Science and Technology, Clear Water Bay, Hong Kong SAR, China

[4] Department of Physics, South University of Science and Technology of China, Shenzhen, Guangdong, China

* To whom correspondence should be addressed. E-mail: phiksou@ust.hk

† These authors contributed equally to this work.

**Superconductivity at the interface of a heterostructure confined to nanometer-sized scale offers unique opportunities to study the exotic physics of two-dimensional superconductivity. The realization of superconductivity at the interface between a topological insulator and an iron-chalcogenide compound is highly attractive for exploring several recent theoretical predictions involving these two new classes of materials. Here, we report transport measurements on a $Bi_2Te_3$/FeTe heterostructure fabricated via van der Waals epitaxy, which demonstrate superconductivity at the interface induced by the $Bi_2Te_3$ epilayer with thickness even down to one quintuple layer. The two-dimensional nature of the observed superconductivity with the highest transition temperature around 12 K was verified by the existence of a Berezinsky-Kosterlitz-Thouless transition and the diverging ratio of in-plane to out-plane upper critical field on approaching the superconducting transition temperature. With the combination of interface superconductivity and Dirac surface states of $Bi_2Te_3$, the heterostructure studied in this work provides a novel platform for realizing Majorana fermions.**

The emergent realization of superconductivity at the interfaces between transition metal oxides (TMOs) has attracted growing attention in recent years[1]. Typical TMO interfaces reported so far include $LaAlO_3$/$SrTiO_3$ (LAO/STO), $LaTiO_3$/$SrTiO_3$, $La_2CuO_4$/$La_{2-x}Sr_xCuO_4$ and $Ba_{1-x}Nd_xCuO_2$/$CaCuO_2$[2-7]. Interface superconductivity between chalcogenides such as PbTe/PbSe and PbSe/PbS has also been reported[8].

Recently, the discovery of superconductivity in layered iron-based materials has steered the high-transition-temperature ($T_c$) superconductors to an iron age. These so-called iron-based superconductors (Fe-SC) are the focus of contemporary research due to their high upper critical fields, large critical current densities and unconventional pairing symmetries. Among them, the 11-type iron-chalcogenide Fe-SCs have the simplest chemical composition and a relatively simple crystalline structure[9-11].

Since the discovery of Fe-SC in 2007, another novel state of quantum matter, the topological insulator (TI), has become an important topic in condensed matter physics[12,13]. The robust topological surface states are topologically protected against time-reversal-invariant perturbations. More interestingly, inducing *s*-wave superconductivity on the surface state of a TI results in a *p*-wave superconductor that hosts Majorana fermions in its vortex cores[12,13]. Motivated by the prospect of creating Majorana fermions, which are non-Abelian particles and have potential applications in quantum computations[14], several attempts have been made to induce superconductivity on TI surfaces using bulk superconductors[15-18]. In this work, we find that interface superconductivity is induced when a thin film of TI, $Bi_2Te_3$, is grown on a non-superconducting FeTe thin film. We show that the resulting superconductivity possesses the characteristics of Fe-SCs with the highest $T_c$ of ~ 12 K. The superconductivity exhibits 2D nature by showing the evidence of a Berezinsky-Kosterlitz-Thouless (BKT) transition and the diverging ratio of in-plane to out-plane upper critical field on approaching $T_c$. Its superconducting thickness is estimated to be ~ 7.0±1.1 nm, based on the measured upper critical fields. This work provides the first example of interface superconductivity between a TI and iron chalcogenide. Due to the possible unconventional pairing symmetry of Fe-SC[19], the $Bi_2Te_3$/FeTe heterostructures also provide new platforms for studying the interplay between unconventional superconductors and TI surface states as well as for creating Majorana fermions.

**Results**

$Bi_2Te_3$ is one of the simplest three-dimensional TIs, whose surface states consist of a single Dirac cone at the $\Gamma$ point[20]. Its unit cell is formed by 3 quintuple layers (QLs), bonded by van der Waals (vdW) forces along the [0001] direction. Each QL contains 2 atomic sheets of Bi and 3 atomic sheets of Te, which are covalently bonded. The unit cell of tetragonal FeTe is composed of two Te - Fe - Te triple layers (TLs) stacked along the [001] direction. Also, the neighboring TLs are bonded via vdW force. Attributed to the vdW bonding nature, even though $Bi_2Te_3$ possesses a six-fold symmetry and FeTe is in a four-fold symmetry with respect to their c-axis directions, the heteroepitaxial growth can still proceed, resulting an atomically sharp interface via vdW expitaxy that allows defect-free epitaxial growth on a substrate possessing large lattice mismatch or distinct crystalline structure as compared with those of the epilayers[21]. Figure 1 displays the cross-section images of a $Bi_2Te_3$ (7 QLs) /FeTe (140 nm) heterostructure taken by a spherical aberration-corrected scanning transmission electron microscope (STEM). Both high-angle annular dark field (HAADF) and annular bright field (ABF) images, as shown in Fig. 1a and b respectively, possess high contrast with significant sensitivity to the atomic number Z, and thus provide robust visibility of both light and heavy atoms. Figure 1c shows the high-magnification HAADF image of this heterostructure, in which one can see that the two layers are separated by a vdW gap and form their own lattices independently, confirming the growth is indeed via vdW epitaxy.

It is well known that FeTe, as the parent compound of iron chalcogenides, does not exhibit superconductivity either at ambient pressure or under high hydrostatic pressure[22-26]. Figure 2a shows the resistance versus temperature $R(T)$ curves for a pure 140-nm-thick FeTe film and a set of $Bi_2Te_3$/FeTe heterostructure samples with different thicknesses of $Bi_2Te_3$ while the thickness of the FeTe film is fixed at 140 nm. All these curves display a common drop around 50 K, which is attributed to the antiferromagnetic transition of FeTe[22]. The pure FeTe film does not show superconductivity down to 2 K. To further confirm the non-superconducting nature of the as-grown FeTe film, two additional FeTe (140 nm) samples were capped with Pd (3 nm) and ZnSe (5 nm), respectively, and in fact their $R(T)$ curves, which are displayed in Fig. 2a as well, neither exhibit any superconductivity. However, the $Bi_2Te_3$/FeTe heterostructure samples with 3, 5, 7, 9 and 14 QLs of $Bi_2Te_3$ undergo a transition into a state with zero resistance. The two heterostructure samples with 1 and 2 QLs of $Bi_2Te_3$ also exhibit a resistance drop although the zero-resistance state has not been reached at the lowest temperature of 280 mK.

The electronic transport data of the heterostructure sample with 7 QLs of $Bi_2Te_3$ (named as Sample A) obtained via standard 4-probe method is displayed in Fig. 2b. One can see that the current density-dependent voltage $V(j)$ of this sample shows a step-like critical current density profile at temperature below 11.0 K. This characteristic $V(j)$ dependence together with the occurrence of a zero-resistance state provide unambiguous evidences for the existence of superconductivity in the $Bi_2Te_3$/FeTe heterostructure. The temperature-dependent critical current density ($j_c$) (defined as the current density at the maximum derivative resistance) as shown in the inset of Fig. 2b enjoys a large magnitude of $10^{-1}$ A/cm. Supplementary Figs. 1a and b, respectively, show that the superconductivity of this heterostructure sample cannot be fully destroyed by applying magnetic fields up to 14 T either vertical ($H_\perp$) or parallel ($H_{//}$) to the interface, which implies its upper critical field ($H_{C2}$) is quite large. The anomalous high $j_c$ and $H_{C2}$ of the $Bi_2Te_3$/FeTe heterostructure sample (similar to those observed in Fe-SC) indicate that its superconductivity may be associated with the induced superconductivity of FeTe (to be discussed later). Furthermore, the superconducting transitions shown in Supplementary Figs. 1a and b display a large anisotropy regarding the direction of the applied magnetic field since the transitions significantly broaden as $H_{C2}^{\perp}$ increases, while such a broadening is much weaker as $H_{C2}^{//}$ increases.

The important role of the $Bi_2Te_3$ thin film for inducing the observed superconductivity in the $Bi_2Te_3$/FeTe heterostructure can be well supported by the results of transport measurements performed on a sample with a 140-nm-thick FeTe film capped with 7 QLs of $Bi_2Te_3$ on half of the FeTe surface, named as Sample B. The detailed characteristics of Sample B addressed in Supplementary Information demonstrate that the $Bi_2Te_3$ thin film is indispensable for the observed superconductivity in the $Bi_2Te_3$/FeTe heterostructure.

It is worth to point out again the fact that all the $Bi_2Te_3$/FeTe heterostructures studied in this work exhibit superconductivity of which the thicknesses of their $Bi_2Te_3$ thin films vary from 1 to 14 QLs (equivalent to ~ 1 to 14 nm), while pure $Bi_2Te_3$ and FeTe films are not superconducting. This fact together with the indispensability of $Bi_2Te_3$ in the observed superconductivity, as shown via Sample B, implies that the observed superconductivity originates from the interfaces of these heterostructures, suggesting that the source of the superconductivity may be confined to two dimensions (2D). The confirmation of the 2D nature of the observed superconductivity can be obtained by testing whether the transport properties of the $Bi_2Te_3$/FeTe heterostructure possess a signature of a BKT transition that is well-known to be characterized by a BKT temperature ($T_{BKT}$), below which a phase transition leading to a 2D topological order emerges[3,6,27-29]. Figure 3a displays the *I-V* isotherms of Sample A on a log-log scale. The straight lines in this plot indicate the power law behaviors, with the powers *α* equal to the slopes of the lines ($V \propto I^{\alpha}$). Among them, the grey line with a slope of 1 coincides with the high-temperature isotherms (11.5 ~ 15.0 K), which indicates their ohmic characteristics ($V \propto I$), while the long black line with a slope of 3 marks the initiation of a BKT transition ($V \propto I^3$). The short black lines are power-law fits to the data at different temperatures. It is worth pointing out that the standard BKT fitting is carried out for data points with current values less than the critical current values corresponding to different temperatures. However, when currents are smaller than some certain values, the system becomes ohmic due to the finite-size effects (FSE)[30]. Thus, the fitting is usually applied to upper portion of the dataset, just below the critical currents. The slopes of the short black lines represent the powers $\alpha$, which are plotted *vs* temperature in Fig. 3b. In this figure, the value of $\alpha$ approaches 3 at temperature of ~ 10.1 K, which is thus identified as $T_{BKT}$, and increases rapidly for temperatures lower than 10.1 K. These observations are regarded as the hallmark of a BKT transition. On the other hand, at the temperature range just above $T_{BKT}$, the temperature-dependent resistance is predicted to be in a form of $R(T) \propto R_0 \exp[-b/(T-T_{BKT})^{1/2}]$, where $R_0$ and $b$ are material-specific parameters[3,6,27-29]. Both the plot of $[d(\ln R)/dT]^{-2/3}$ versus $T$ as shown in Fig. 3c and the best fit to the $R(T)$ data using this form as shown in Fig. 3d agree well with the expected BKT behavior near the transition, yielding $T_{BKT}$ equal to 10.2 and 9.9 K, respectively. These two resulting $T_{BKT}$ values are highly consistent with the value extracted from the power analysis, 10.1 K. Our analysis thus provides strong evidences for a 2D nature of the observed superconductivity.

Figure 4 shows the temperature-dependent upper critical field $H_{C2}(T)$ for Sample A in directions parallel ($H_{C2}^{//}$) and perpendicular ($H_{C2}^{\perp}$) to the interface. The zero-resistance temperature (corresponding to the temperature at which the zero-resistance state is reached) is extracted from the $R(T)$ characteristics under different $H_{\perp}$ and $H_{//}$ displayed in Supplementary Figs. 1a and b. The corresponding $H_{C2}(T)$ values derived from the zero-resistance temperatures were used to estimate the Ginzburg-Landau (GL) coherence length $\xi_{GL}$ and superconducting thickness $d_{sc}$ in our $Bi_2Te_3$(7 QLs)/FeTe heterostructure. It is noted that upper critical field for the parallel field direction $H_{C2}^{//}$ follows the GL form of temperature-dependent behavior for a 2D superconducting film, $H_{C2}^{//}(T) = \frac{\sqrt{3}\Phi_0}{\pi\xi_{GL}(0)d_{sc}}\left(1-\frac{T}{T_C}\right)^{1/2} \propto \left(1-\frac{T}{T_C}\right)^{1/2}$, where $\Phi_0$ is the flux quantum and $\xi_{GL}(0)$ is the GL coherence length at $T=0$ K, and the resulting fitting curve is shown in Fig. 4 as well. In the perpendicular field direction, $H_{C2}^{\perp}$ shows the expected linear $T$ dependence, which follows the standard linearized GL theory for 2D superconductors, $H_{C2}^{\perp} = \frac{\Phi_0}{2\pi\xi_{GL}(0)^2}\left(1-\frac{T}{T_C}\right) \propto \left(1-\frac{T}{T_C}\right)$. The derived $H_{C2}(T)$ values were used to plot the $H_{C2}^{//}/H_{C2}^{\perp}$ ratio versus reduced temperature $T/T_c$ (inset of Fig. 4) and it shows a diverging characteristic on approaching $T_c$, characteristic for a 2D nature[31]. Using this temperature-dependent $H_{C2}^{\perp}$ relationship, the mean value of $\xi_{GL}(0)$ can be calculated to be 5.2±1.7 nm. The mean value of superconducting thickness $d_{sc}$ is then estimated to be 7.0±1.1 nm using $d_{sc} = \frac{\sqrt{3}\Phi_0}{\pi\xi_{GL}(0)H_{C2}^{//}(T)}\left(1-\frac{T}{T_C}\right)^{1/2}$ based on the extracted $H_{C2}^{//}(T)$ values and fittings. The fact that the zero-temperature critical field of FeTe is much larger than the fields used in taking the data for performing the fitting will certainly cause a substantial uncertainty in the estimations of both $\xi_{GL}$ and $d_{sc}$. As their fitted values are quite close, one can still regard as a further confirmation for the 2D nature of the observed superconductivity, which is mainly demonstrated by the good match of the transport data with the BKT model and the diverging ratio of $H_{C2}^{//}/H_{C2}^{\perp}$ on approaching $T_c$. It is worth pointing out that in parallel applied fields, the upper critical field (orbital limit for superconductivity) may be replaced by the Pauli limit for superconductivity. However, in the derivation of the coherence length the extrapolated orbital limit must be used, which agrees with our method of extrapolating the critical field line from our data at low fields close to the zero field critical temperature.

In the following, we would perform the study regarding the crucial role played by FSE and inhomogeneity on the BKT transition in our $Bi_2Te_3$(7 QLs)/FeTe heterostructure as well as discuss the validity of our BKT analysis. Firstly, in the infinite-size homogeneous case, the best fit using the infinite-size formula $\frac{R}{R_0} \simeq 10.8 b \exp(-2\sqrt{bt_c/t})$ in Ref. 32 (where $t = \frac{T - T_{BKT}}{T_{BKT}}$ and $t_c = \frac{T_c - T_{BKT}}{T_{BKT}}$ are the reduced temperatures) for the $R$-$T$ data of our $Bi_2Te_3$(7 QLs)/FeTe sample is obtained with fitted parameters of $b\sim0.98$, $T_{BKT}\sim10.1$ K, and mean-field $T_c\sim11.1$ K. Thus the range between the mean-field $T_c$ and $T_{BKT}$ covers from 11.1 to 10.1 K, which is consistent with the temperature range used in the fittings presented in Fig. 3c and d. The value of the parameter "$b$" in the equation of $R(T) \propto R_0\exp[-b/(T-T_{BKT})^{1/2}]$ mentioned above is determined to be ~2.1 in fitting the standard BKT model. The fact that "$b$" is in the order of 1, is consistent with the expectation of a recent theoretical model[33]. Here, "$b$" was mentioned as a material parameter in the related works regarding 2D superconductors[3,6]. As pointed out in Ref. 33, "$b$" appears in the exponential expression for the BKT correlation length $\frac{\xi}{\xi_0} = \frac{1}{A}\exp(b/\sqrt{t}), t \to 0$. "$b$" is meaningful in the sense that it appears in the condition $b/\sqrt{t} \gg 1$ for the use of the correlation length equation mentioned above. In our case, $b/\sqrt{t} \sim 6.7$ at the lower limit when $t = t_c = \frac{T_c - T_{BKT}}{T_{BKT}}$, thus, well satisfies the condition.

In addition, the rounding of the transition near $R=0$ in the $R$-$T$ curve shown in Fig. 2a is expected to be due to FSE. Based on Ref. 33, we divided the $R$-$T$ transition into four regions, GL-fluctuation region, BKT transition region, FSE-dominated region, and superconducting region, respectively, as shown in Fig. 5, where the data are well fitted with the interpolating formula in Ref. 33 for an inhomogeneity effect model (red curve). At $T<T_{BKT}$, FSE arises and the normalized resistance starts to deviate from the infinite-size limit (black curve). As described in Ref. 33, inhomogeneity effects (or FSE) is associated with the scale of the sample dimensions or even smaller, and even though the system does not have a true granular structure, the homogeneous regions will have a typical size $L_{hom}$ smaller than the physical size $L_{phys}$ of the sample. The physical dimension of our sample is $L_{phys} \sim 500$ μm and the critical current at $T_{BKT}$ is $I_c \sim 5\times10^{-3}$ A. Using the equation $I_c \simeq 4K_B T_{BKT} \frac{c}{\Phi_0} \frac{L_{phys}}{L_{hom}}$ shown in Ref. 33, the characteristic length scale of our sample is estimated to be $L_{hom} \sim 0.3$ μm. As expected by Ref. 33, $L_{hom} < L_{phys}$ will give rise to a critical current for linear-to-nonlinear characteristic larger than expected for the homogeneous case, which is also consistent with the characteristics of $j_c$ vs $T$ relationship of our sample as shown in the inset of Fig. 2b. In the infinite-size homogeneous case, a jump is expected from $\alpha=3$ to $\alpha=1$, however, for both our heterostructure sample and the LAO/STO interface superconductor[3], the $\alpha$ value seems to show a smooth transition rather than a sudden sharp jump. As explained in Ref. 33, when taking the FSE into consideration, the change of $\alpha$ will be observed to possess a smooth downturn near $T_{BKT}$, which is exactly what we observed in our data.

**Discussion**

To reveal the origin of the 2D superconductivity at the interface of the heterostructure through either a theoretical or an experimental approach always remains as a big challenge. Based on the data extracted from Fig. 2a, one can relate $T_{c(onset)}$ of the $Bi_2Te_3$/FeTe heterostructures with the $Bi_2Te_3$-thickness. As shown in Fig. 2c, $T_{c(onset)}$ increases steadily with the increase of the thickness of $Bi_2Te_3$, and then reaches a plateau value of ~ 11.5 K when the thickness of $Bi_2Te_3$ reaches 5 QLs, a critical thickness that sets the length scale of $Bi_2Te_3$ for observing the 2D superconductivity. Such dependence seems to suggest that chemical doping of FeTe, perhaps involving Bi, may be the cause of the observed superconductivity. This possible mechanism was investigated by carrying out atomic-resolution energy-dispersive X-ray spectroscopy (EDS) mapping across the $Bi_2Te_3$/FeTe interface as shown in Fig. 1d, which however shows that the fall of the Bi profile at the interface is quite sharp within the spatial resolution limit of atomic-resolution EDS derived from atomically sharp interfaces[34]. We have also performed extrinsic Bi doping of FeTe in two approaches, one with an elemental Bi source and the other with a $Bi_2Te_3$ compound source. It was found that the *R vs T* behaviors of these samples do not show any superconductivity feature (Supplementary Fig. 3). Both the results of atomic-resolution EDS and extrinsic Bi doping described above seem to imply that the observed 2D superconductivity at the $Bi_2Te_3$/FeTe interface may not be caused by Bi incorporation in the FeTe layer. We also provide a discussion about possible intrinsic Te doping in FeTe layer in the Supplementary Information, which also implies that intrinsic Te doping is also unlikely to be the cause of the observed superconductivity.

It is also worth to point out that one may argue this superconductivity may come from strain-induced superconductivity in FeTe as observed when FeTe was grown on lattice-mismatch substrates[35]. However, the growth of the top $Bi_2Te_3$ layer on the bottom FeTe layer of the heterostructures studied in this work is via vdW epitaxy that is believed not to introduce substantial lattice distortion on both layer components. This is further confirmed by the results of a statistical approach used to determine the lattice parameters of FeTe and $Bi_2Te_3$. Figure 1e shows the variation of the lattice parameters of FeTe (left) and $Bi_2Te_3$ (right) extracted from their HAADF images. As can be seen in this figure, except a small crystalline distortion appears in a portion of FeTe near the FeTe/ZnSe interface, the lattice parameters ($a$ and $c$) of both the $Bi_2Te_3$ and FeTe portions away from the $Bi_2Te_3$/FeTe interface well match with their bulk values. Thus, significant strain built up at the interface of the heterostructure is unlikely.

As discussed above, the anomalous high $j_c$ and $H_{C2}$ of the $Bi_2Te_3$/FeTe heterostructure sample are similar to those observed in Se doped FeTe thin films, in which the antiferromagnetic order of the parent FeTe is destroyed and the sample becomes superconducting. This indicates that the superconductivity of the $Bi_2Te_3$/FeTe heterostructure may be associated with the appearance of a superconducting layer of FeTe near the interface. It is possible that the significantly n-doped $Bi_2Te_3$ thin film, with the presence of the topological surface states, may increase the electron density of the FeTe layers near the interface. This change of electron density may destroy the antiferromagnetic order of FeTe[22] and turn it into a superconductor. The $Bi_2Te_3$ thin film fabricated in our MBE system has been previously characterized to be significantly n-doped with the Fermi level lying above the bottom of the conduction band[36]. As shown by Li *et al.*[37], through performing ARPES (angle-resolved photoemission spectroscopy) on MBE-grown $Bi_2Te_3$ thin films, the surface states can even be detected for thin films down to a single QL of $Bi_2Te_3$ even though an energy gap opens near the Dirac point due to the coupling between the top and bottom surface states. Within this energy gap, the surface states do not exist as expected. However, the surface states still exist at chemical potential well above the Dirac point, for example, when the sample is heavily n-doped. Therefore, observing reasonably strong superconductivity for heterostructures with $Bi_2Te_3$ thin film down to a single QL does not rule out that surface states may play a role in the emergence of superconductivity in these samples. Moreover, Li *et al.*'s work also reveals that as the thickness of $Bi_2Te_3$ increases towards 5 QLs, the surface states reach their full topological nature, which means 5 QLs sets the length scale of fully-developed surface states for $Bi_2Te_3$. In our case, as shown in Fig. 2c, 5-QL is a critical thickness that sets the length scale of $Bi_2Te_3$ for observing the 2D superconductivity. Since the two length scales mentioned above happen to coincide with each other, it may indicate that the topological nature of $Bi_2Te_3$ and the observed 2D superconductivity perhaps are correlated with each other. However, it should be pointed out that as the Fermi level of our $Bi_2Te_3$ layer lies above the bottom of the conduction band, the roles of the surface states and bulk states of $Bi_2Te_3$ cannot be distinguished. Further studies, such as realizing an intrinsic $Bi_2Te_3$ layer or tuning the Fermi level to coincide with the Dirac point through either extrinsic doping or adjusting gate bias, may help to clarify this issue.

Below the novelty and importance of this work are addressed. As compared with the superconductivities found in oxide heterostructures and arrays of superconducting islands on gold[38], our work enjoys several novelties: firstly, an important novelty lies on the difference in the constituents of the materials. The interface superconductivity of our $Bi_2Te_3$/FeTe heterostructure likely requires the delicate interplay between the non-superconducting TI $Bi_2Te_3$ and the non-superconducting parent compound of Fe-based superconductor FeTe. Our heterostructure system forms a novel platform for studying the interactions between these two new quantum states of matters. On the other hand, the superconductivity at the interface of oxide heterostructures (like LAO/STO) arises from the interplay of two traditional oxides. Moreover, the optimal $T_c$ of our system is about 12 K which is 50 times higher than the $T_c$ found in LAO/STO. Secondly, to the best of our knowledge, our $Bi_2Te_3$/FeTe heterostructure and the one unit-cell-thick FeSe grown on SrTiO$_3$ substrate[9,11] are the only 2D superconductivity systems, reported so far, that consist of an iron chalcogenide component. The 2D superconductivity of our heterostructure distinguishes from that of the latter system by the fact that it arises from an addition of a new component (Bi$_2$Te$_3$) while the latter simply relies on its own dimensionality reduction. Thirdly, comparing with the other 2D superconducting systems reported so far, the Bi$_2$Te$_3$ layer in our heterostructure enjoys a strong intrinsic spin-orbit coupling[20] while the inversion symmetry of the observed 2D superconductivity is broken at the interface, this combination could give rise to Rashba-type spin-orbit interactions and cause unconventional superconducting paring symmetry at the interface, similar to the LAO/STO system in which Rashba spin-orbit coupling is important for the formation of the helical Fulde-Ferrell-Larkin-Ovchinnikov (FFLO) states[39]. Fourthly, our work provides the first example of interface superconductivity between a TI and iron chalcogenide. Due to the possible unconventional pairing symmetry of Fe-based superconductor[19], the Bi$_2$Te$_3$/FeTe heterostructures also provide new platforms for creating and studying Majorana fermions in a 2D template, while several previous attempts have been made to induce superconductivity on TIs using bulk traditional superconductors.

In summary, a Bi$_2$Te$_3$/FeTe heterostructure, of which both components are not intrinsic superconductors, was demonstrated to possess 2D superconductivity at the interface, which is found to be induced by the Bi$_2$Te$_3$ epilayer. The existence of a BKT transition supported by a detailed analysis of the transport properties and the diverging ratio of in-plane to out-plane upper critical field on approaching $T_c$ together form solid evidences of the 2D nature of the observed superconductivity. We have discussed the underlying mechanism of the 2D superconductivity of the heterostructure through analyzing the strain behavior and possible inter-diffusion across the interface as well as the $Bi_2Te_3$-thickness-dependent critical temperature. We suggest that the heavily n-doped $Bi_2Te_3$ layer with topological surfaces may increase the charge density of the FeTe layers near the interface and induce superconductivity. The $Bi_2Te_3$/FeTe heterostructure also provides a novel platform for the study of Majorana fermions due to the presence of interface superconductivity and the topological surface states of $Bi_2Te_3$.

**Methods**

**Film Growth.** All the samples studied in this work were synthesized by a VG-V80H MBE system. Each sample contains a ZnSe buffer layer (50 nm) firstly deposited on the GaAs (001) semi-insulating substrates, followed by a deposition of FeTe with a thickness of 140 nm. The $Bi_2Te_3$/FeTe heterostructure samples were fabricated with varying thicknesses of the $Bi_2Te_3$ thin films from 1, 2, 3, 5, 7, 9 and 14 QLs. A special heterostructure sample with $Bi_2Te_3$ (7 QLs) deposited on only half of the FeTe (140 nm) surface was prepared by partially covering the FeTe surface with a tantalum strip *in situ* prior to the deposition of $Bi_2Te_3$.

**Device Fabrications and Transport Measurements.** Each sample was cut into long strips using diamond scriber. Silver paint was used to form circular electrodes with diameters of ~ 0.5 mm and spacing of ~ 2.0 mm, which were connected to the measuring instruments with aluminum wires. Silver has long been characterized as a fast diffuser in $Bi_2Te_3$[40,41]. Based on our calculation using these findings in this reference, it takes the silver paint just takes 7 minutes to diffuse to a depth of 100 nm in $Bi_2Te_3$. Thus, it is likely that our silver paint electric contacts should penetrate the entire $Bi_2Te_3$ thin film and reach the interface of the heterostructure and beyond. All the transport measurements were carried out in a Quantum Design PPMS system, which is equipped with a 14-Tesla superconducting magnet and possesses a base temperature of 2 K. The transport measurements were conducted using a pulse mode to avoid heating effect, which indeed was found to be negligible for our measurements as consistent results were obtained either the samples were placed in Helium-4 vapor or in Helium-4 liquid. For the heterostructure samples with thicknesses of 1, 2 and 3 QLs of $Bi_2Te_3$, the measuring temperature was further cooled down to 280 mK in an Oxford Cryostat with Helium-3 insert.

**References**

1   Hwang, H. Y. *et al.* Emergent phenomena at oxide interfaces. *Nature Mater.* **11**, 103-113 (2012).

2   Balestrino, G. *et al.* Very large purely intralayer critical current density in ultrathin cuprate artificial structures. *Phys. Rev. Lett.* **89**, 156402 (2002).

3   Reyren, N. *et al.* Superconducting interfaces between insulating oxides. *Science* **317**, 1196-1199 (2007).

4   Orgiani, P. *et al.* Direct measurement of sheet resistance $R_\Box$ in cuprate systems: evidence of a fermionic scenario in a metal-insulator transition. *Phys. Rev. Lett.* **98**, 036401 (2007).

5   Gozar, A. *et al.* High-temperature interface superconductivity between metallic and insulating copper oxides. *Nature* **455**, 782-785 (2008).

6   Yuli, O. *et al.* Enhancement of the superconducting transition temperature of $La_{2-x}Sr_xCuO_4$ bilayers: role of pairing and phase stiffness. *Phys. Rev. Lett.* **101**, 057005 (2008).

7   Biscaras, J. *et al.* Two-dimensional superconductivity at a Mott insulator/band insulator interface $LaTiO_3/SrTiO_3$. *Nature Commun.* **1**, 89 (2010).

8   Fogel, N. *et al.* Direct evidence for interfacial superconductivity in two-layer semiconducting heterostructures. *Phys. Rev. B* **73**, 161306(R) (2006).

9   Wang, Q. Y. *et al.* Interface-induced high-temperature superconductivity in single unit-cell FeSe films on $SrTiO_3$. *Chin. Phys. Lett.* **29**, 037402 (2012).

10  Liu, D. *et al.* Electronic origin of high-temperature superconductivity in single-layer FeSe superconductor. *Nature Commun.* **3**, 931 (2012).

11  Tan, S. *et al.* Interface-induced superconductivity and strain-dependent spin density waves in $FeSe/SrTiO_3$ thin films. *Nature Mater.* **12**, 634-640 (2013).

12  Fu, L. & Kane, C. Superconducting proximity effect and Majorana fermions at the surface of a topological insulator. *Phys. Rev. Lett.* **100**, 096407 (2008).

13   Qi, X. L. & Zhang, S. C. Topological insulators and superconductors. *Rev. of Mod. Phys.* **83**, 1057-1110 (2011).

14   Nayak, C., Stern, A., Freedman, M. & Das Sarma, S. Non-Abelian anyons and topological quantum computation. *Rev. of Mod. Phys.* **80**, 1083 (2008).

15   Williams, J. R. *et al.* Unconventional Josephson effect in hybrid superconductor-topological insulator devices. *Phys. Rev. Lett.* **109**, 056803 (2012).

16   Qu, F. *et al.* Strong superconducting proximity effect in Pb-$Bi_2Te_3$ hybrid structures. *Sci. Rep.* **2**, 339 (2012).

17   Veldhorst, M. *et al.* Josephson supercurrent through a topological insulator surface state. *Nature Mater.* **11**, 417-421 (2012).

18   Wang, M. X. *et al.* The coexistence of superconductivity and topological order in the $Bi_2Se_3$ thin films. *Science* **336**, 52 (2012).

19   Hanaguri, T., Niitaka, S., Kuroki, K. & Takagi, H. Unconventional *s*-wave superconductivity in Fe(Se,Te). *Science* **328**, 474-476 (2010).

20   Chen, Y. L. *et al.* Experimental realization of a three-dimensional topological insulator, $Bi_2Te_3$. *Science* **325**, 178-181 (2009).

21   Geim, A. K. & Grigorieva, I. V. Van der Waals heterostructures. *Nature* **499**, 419-425 (2013).

22   Chen, G. F. *et al.* Electronic properties of single-crystalline $Fe_{1.05}Te$ and $Fe_{1.03}Se_{0.30}Te_{0.70}$. *Phys. Rev. B* **79**, 140509(R) (2009).

23   Li, S. *et al.* First-order magnetic and structural phase transitions in $Fe_{1+y}Se_xTe_{1-x}$. *Phys. Rev. B* **79**, 054503 (2009).

24   Bao, W. *et al.* Tunable (δπ, δπ)-type antiferromagnetic order in α-Fe(Te,Se) superconductors. *Phys. Rev. Lett.* **102**, 247001 (2009).

25   Mizuguchi, Y., Tomioka, F., Tsuda, S., Yamaguchi, T. & Takano, Y. FeTe as a candidate material for new iron-based superconductor. *Physica C: Superconductivity* **469**, 1027-1029 (2009).

26   Zhang, C. *et al.* Pressure-induced lattice collapse in the tetragonal phase of single-crystalline $Fe_{1.05}Te$. *Phys. Rev. B* **80**, 144519 (2009).

27   Kosterlitz, J. M. & Thouless, D. J. Ordering, metastability and phase transitions in two-dimensional systems. *J. Phys. C: Solid State Phys.* **6**, 1181-1203 (1973).

28   Kosterlitz, J. M. The critical properties of the two-dimensional xy model. *J. Phys. C: Solid State Phys.* **7**, 1046-1060 (1974).

29   Halperin, B. I. & R., N. D. Resistive transition in superconducting films. *J. of Low Temp. Phys.* **36**, 599-616 (1979).

30   Medvedyeva, K., Kim, B. J. & Minnhagen, P. Analysis of current-voltage characteristics of two-dimensional superconductors Finite-size scaling behavior in the vicinity of the Kosterlitz-Thouless transition. *Phys. Rev. B* **62**, 14531-14540 (2000).

31   Y. Mizukami *et al.* Extremely strong-coupling superconductivity in artificial two-dimensional Kondo lattices. *Nature Phys.* **7**, 849-853 (2011).

32   Hsu, J. & Kapitulnik, A. Superconducting transition, fluctuation, and vortex motion in a two-dimensional single-crystal Nb film. *Phys. Rev. B* **45**, 4819-4835 (1992).

33   Benfatto, L., Castellani, C. & Giamarchi, T. Broadening of the Berezinskii-Kosterlitz-Thouless superconducting transition by inhomogeneity and finite-size effects. *Phys. Rev. B* **80**, 214506 (2009).

34   Williams, D. B. & Carter, C. B. Transmission electron microscopy. 663-677, 2nd ed. Springer (2009).

35   Han, Y. *et al.* Superconductivity in iron telluride thin films under tensile stress. *Phys. Rev. Lett.* **104**, 017003 (2010).

36   He, H.-T. *et al.* Impurity effect on weak antilocalization in the topological insulator $Bi_2Te_3$. *Phys. Rev. Lett.* **106**, 166805 (2011).

37   Li, Y.-Y. *et al.* Intrinsic topological insulator $Bi_2Te_3$ thin films on Si and their thickness limit. *Adv. Mater.* **22**, 4002-4007 (2010).

38   Eley, S., Gopalakrishnan, S., Goldbart, P. & Mason, N. Approaching zero-temperature metallic states in mesoscopic superconductor–normal–superconductor arrays. *Nature. Phys.* **8**, 59-62 (2011).

39   Michaeli, K., Potter, A. C. & Lee, P. A. Superconducting and ferromagnetic phases in $SrTiO_3$/$LaAlO_3$ oxide interface structures: Possibility of finite momentum pairing. *Phys. Rev. Lett.* **108**, 117003 (2012).

40   Dibbs, H. P. Thermal diffusion of silver in single-crystal bismuth telluride. *J. App. Phys.* **39**, 2976-2977 (1968).

41   Keys, J. & Dutton, H. Diffusion and solid solubility of silver in single-crystal bismuth telluride. *J. Phys. Chem. Solids* **24**, 563-571 (1963).

**Acknowledgments:**

We gratefully acknowledge the use of the facilities in the Materials Characterization and Preparation Facility (MCPF) at the Hong Kong University of Science and Technology. We also thank Dr. Emrah Yücelen, the senior research scientist in NanoPort Europe of FEI Company, for performing the STEM experiments and useful discussion, as well as Miss Junying Shen for the help in performing the transport measurements. The work described here was substantially supported by grants from the Research Grants Council of the Hong Kong Special Administrative Region, China (project No. 604910, 605011, AOE/P-04/08-3, 605512 and 603010).

**Author Contributions**

Q.L.H. and H.L. initiated this study; I.K.S. and Q.L.H. further designed the experiments with contributions from H.L., J.W., and R.L.; Q.L.H. carried out the sample growth and structural characterization with contributions from G.W. and Y.H.L.; H.L., H.H., and M.H. contributed the transport measurements; Q.L.H., J.W., R.L., K.T.L., and I.K.S. performed the data analysis; all authors contributed to the scientific planning and discussions; I.K.S. and Q.L.H. wrote the manuscript with contributions from other authors.

**Competing Financial Interests statement**

The authors declare no competing financial interests.

**Figure Legends**

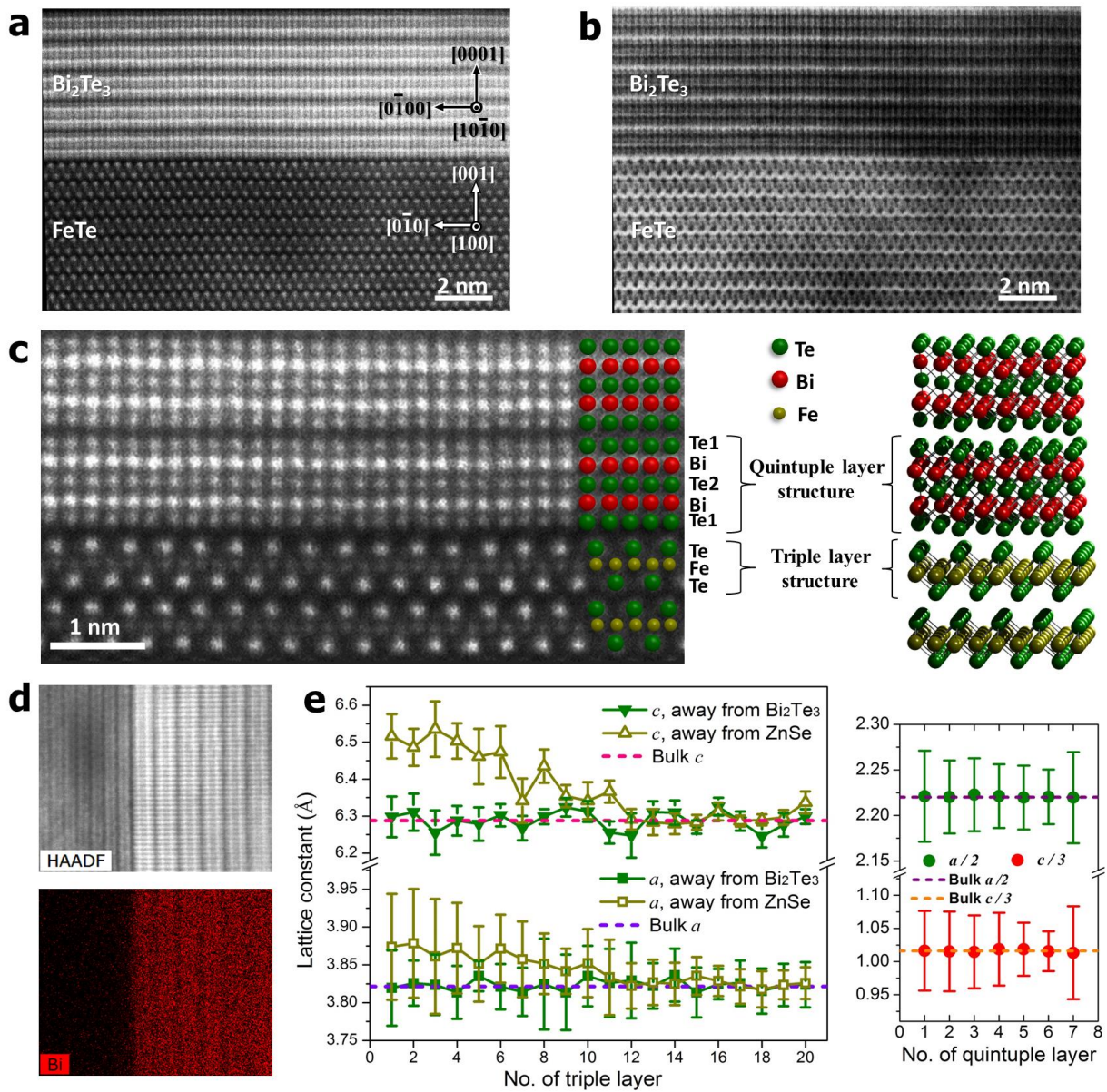

**Figure 1**

**Cross-sectional spherical aberration-corrected scanning TEM micrographs of a Bi$_2$Te$_3$(7 QLs)/FeTe heterostructure.** (**a**) HADDF image. (**b**) ABF image. (**c**) Higher-magnification HADDF image shows atomically sharp interface between Bi$_2$Te$_3$ and FeTe. (**d**), Bi-EDS mapping across the interface. (**e**), Variation of the lattice parameters of FeTe (left) and $Bi_2Te_3$ (right), extracted from their HAADF images.

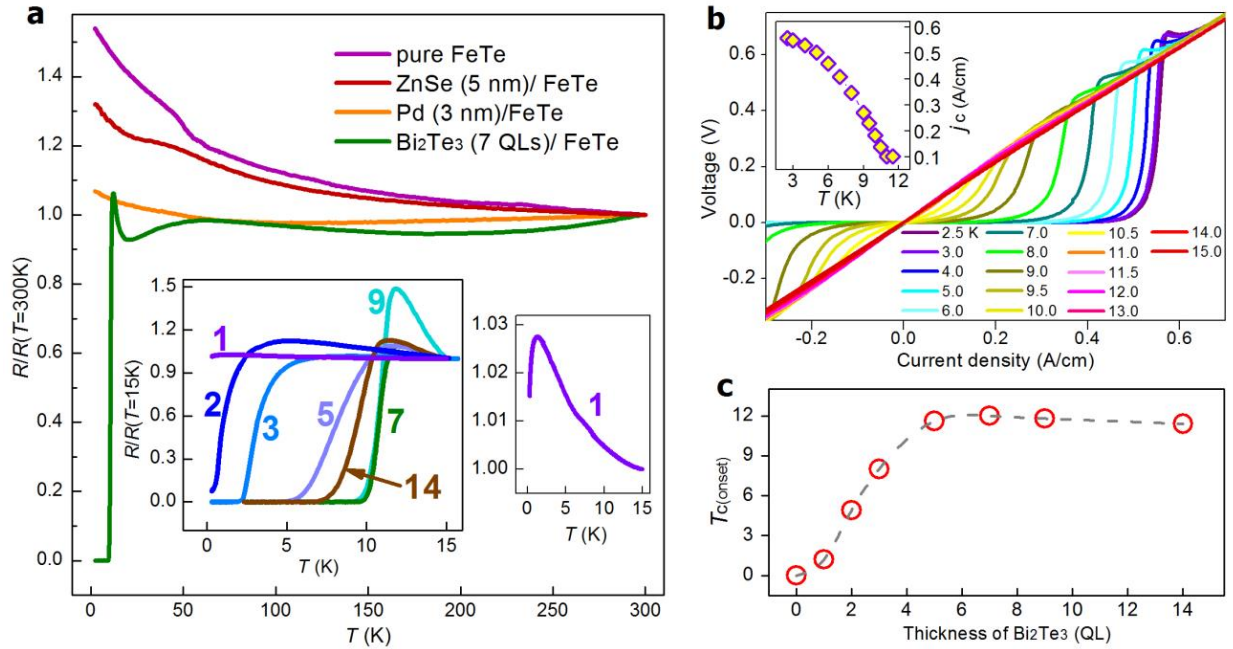

**Figure 2**

**Results of transport measurements.** (**a**) Temperature-dependent resistance $R(T)/R(T=300K)$ of pure FeTe, ZnSe/FeTe and Pd/FeTe films as well as a $Bi_2Te_3$(7 QLs)/FeTe heterostructure. The room temperature resistances of these samples are 16.1, 18.7, 5.0, and 22.5 $\Omega$, respectively. The bottom left inset shows the $R(T)/R(T=15K)$ versus temperature of $Bi_2Te_3$/FeTe heterostructures with $Bi_2Te_3$ thicknesses labeled in units of QL, while the bottom right inset is a finer scale plot for a $Bi_2Te_3$(1 QL)/FeTe heterostructure. (**b**) Current-density-dependent voltage of the $Bi_2Te_3$(7 QLs)/FeTe heterostructure at different temperatures. The inset shows its temperature-dependent critical current density. (**c**) $T_{c(onset)}$ versus the thickness of $Bi_2Te_3$ thin film in units of QL. The dash line is a guide to the eyes.

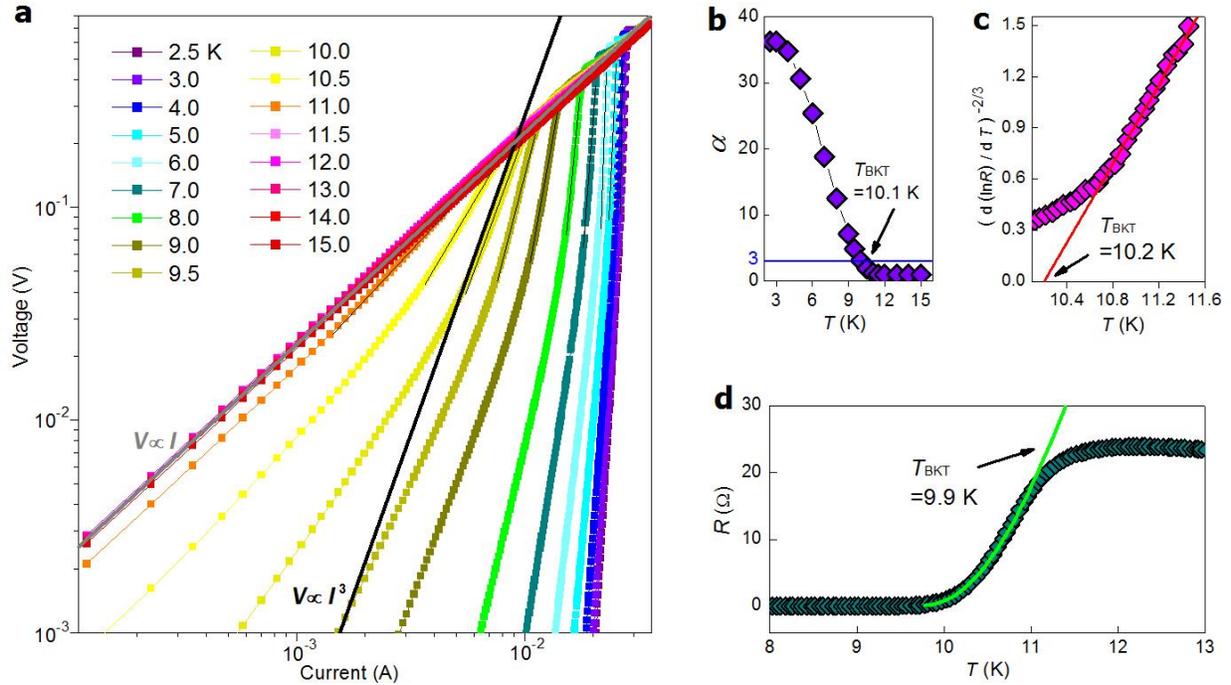

**Figure 3**

**Low-temperature transport properties of Sample A: $Bi_2Te_3$(7 QLs)/FeTe heterostructure.**
(**a**) Voltage-current density data plotted in a log-log scale. The short black lines are power-law fits of the data in the BKT transitions at different temperatures. The grey line corresponds to a $V \propto I$ behavior while the black long line corresponds to a $V \propto I^3$ behavior. (**b**) Temperature dependence of the power-law exponent α deduced from the power-law fits in (**a**). The interception with the solid line $y=3$ gives a $T_{BKT}$=10.1 K. (**c**) $R(T)$ dependence plotted on a $[dln(R)/dT]^{-2/3}$ versus $T$ scale. The solid line is the behavior expected for a BKT transition with $T_{BKT}$=10.2 K. (**d**) $R(T)$ dependence with a fitting for a BKT behavior near the transition, yielding a $T_{BKT}$=9.9 K.

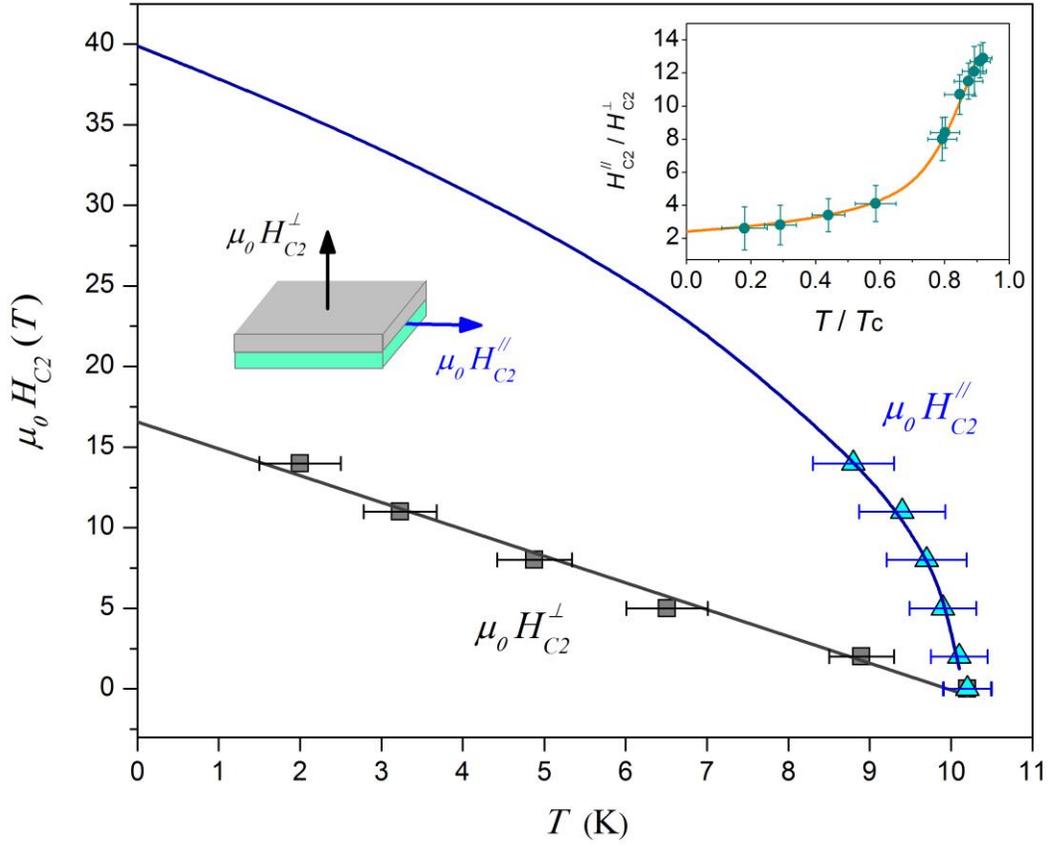

**Figure 4**

**Upper critical fields in the directions parallel and perpendicular to the interface of Bi$_2$Te$_3$/FeTe heterostructure as a function of the temperature.** The curves are resulted from the fittings to the data based on the GL theory for 2D superconductors. Inset: $H_{C2}^{//}/H_{C2}^{\perp}$ as a function of reduced temperature $T/T_c$, showing its diverging nature on approaching $T_c$.

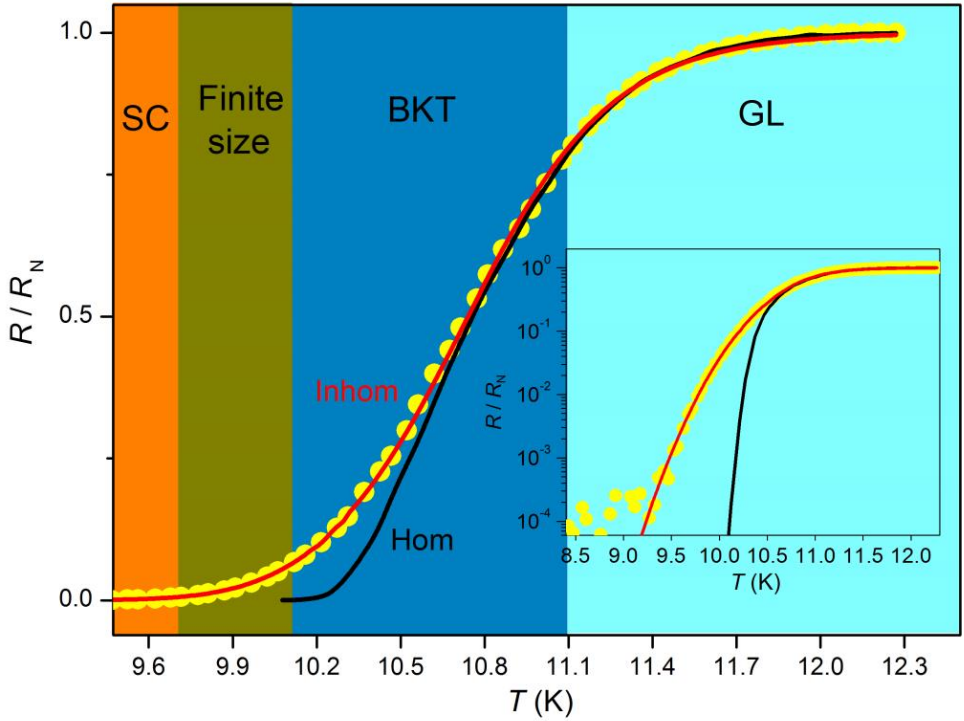

**Figure 5**

**Normalized resistance *vs* temperature of a Bi$_2$Te$_3$(7 QLs)/FeTe heterostructure plotted in four regions covering from GL, BKT, FSE to SC.** The inset shows the same curve in a logarithmic plot. The red curve is a fit with the interpolating formula in Ref. 33 for an inhomogeneity effect model while the black curve is a fit based on the infinite-size limit.

# Supplementary Information for:

# Two-dimensional superconductivity at the interface of a $Bi_2Te_3$/FeTe heterostructure

Qing Lin He[2,3]†, Hongchao Liu[1]†, Mingquan He[1,2], Ying Hoi Lai[2,3], Hongtao He[1,4], Gan Wang[4], Kam Tuen Law[1], Rolf Lortz[1,2], Jiannong Wang[1,2] and Iam Keong Sou[1,2,3]∗

[1] Department of Physics, the Hong Kong University of Science and Technology, Clear Water Bay, Hong Kong SAR, China

[2] William Mong Institute of Nano Science and Technology, the Hong Kong University of Science and Technology, Clear Water Bay, Hong Kong SAR, China

[3] Nano Science and Technology Program, the Hong Kong University of Science and Technology, Clear Water Bay, Hong Kong SAR, China

[4] Department of Physics, South University of Science and Technology of China, Shenzhen, Guangdong, China

∗ To whom correspondence should be addressed. E-mail: phiksou@ust.hk

† These authors contributed equally to this work.

**Supplementary Information:**

Supplementary Text

Supplementary Table 1

Supplementary Figures 1-5

Supplementary References 1-5

**Supplementary Text：**

**The important role of $Bi_2Te_3$ for the observed 2D superconductivity.**

To show the important role of the $Bi_2Te_3$ thin film for inducing the observed superconductivity in the $Bi_2Te_3$/FeTe heterostructure, a sample with a 140-nm-thick FeTe film capped with 7 QLs of $Bi_2Te_3$ on half of the FeTe surface was fabricated (named as Sample B). Sample B was cut into long strips using diamond scriber. Silver paint was used to form circular electrodes with diameters of ~ 0.5 mm and spacing of ~ 2.0 mm, which were connected to the measuring instruments with aluminum wires. Its transport properties were measured in/across the regions with and without $Bi_2Te_3$ using 8 separated electrodes by applying the current and then measuring the voltage as shown in Supplementary Fig. 2a, where the temperature-dependent resistances $R(T)$ were obtained from the $V_0$-$I_0$ measurement via a standard 4-probe method as well as from the $V_0$-$I_0^*$, $V_1$-$I_1$ and $V_2$-$I_2$ measurements via a pseudo 4-probe method. As shown in the resulting $R(T)$ curves of Sample B (Supplementary Fig. 2b), the region fully covered with $Bi_2Te_3$ shows a single superconducting transition, while the region without $Bi_2Te_3$, equivalent to a pure FeTe film, displays a semiconducting behavior without any signature of superconductivity down to the lowest measuring temperature. The resulting $R(T)$ curves from the $V_0$-$I_0$ and $V_0$-$I_0^*$ measurements both reveal the electronic transport properties across the region consisting of a $Bi_2Te_3$/FeTe heterostructure and a pure FeTe film. These two $R(T)$ curves show similar characteristics in their superconducting transitions except different residual resistances. It can be seen that the resistance continues to drop gradually below the main superconducting transition which extends over the temperature range from ~ 6 K to 10 K. This feature can be clearly demonstrated by the enhanced intensity below the peak associated with the main transition in the corresponding d$R$/d$T$ curves shown in the inset of Supplementary Fig. 2b, as marked by a black arrow. It provides evidence for the existence of a superconducting proximity effect in the region consisting of a $Bi_2Te_3$/FeTe heterostructure and a pure FeTe film. It is believed that the first resistance drop is attributed to the superconductivity originated from the $Bi_2Te_3$/FeTe heterostructure, while the second one comes from a proximity-induced superconductivity in a fraction of the pure FeTe film, which is located in the vicinity of the heterostructure (indicated by a red bar in Supplementary Fig. 2a). All the characteristics of Sample B addressed in this paragraph demonstrate that the $Bi_2Te_3$ thin film is indispensable for the observed superconductivity in the $Bi_2Te_3$/FeTe heterostructure.

**Extrinsic Bi doping of FeTe.**

Using an additional elemental Bi source, two multilayer FeTe:Bi samples (named as FeTe:Bi#A and FeTe:Bi#B) with different Bi-source temperatures covering from 265 to 405 ℃ were fabricated. Another multilayer FeTe:Bi sample (named as FeTe:Bi#C) was fabricated using $Bi_2Te_3$ compound as the dopant source with dopant source temperatures covering from 370 to 410 ℃, in which 410℃ is the same source temperature used to grow the $Bi_2Te_3$ layer of the $Bi_2Te_3$/FeTe heterostructures so as to achieve a condition close to that used in the growth of heterostructures. The inset of Supplementary Fig. 3 shows their layer structures and Supplementary Table 1 displays their corresponding Bi/$Bi_2Te_3$-source temperatures. Time-of-flight Secondary Ion Mass Spectrometry (Tof-SIMS, which has a detection limit of $10^{-6} \sim 10^{-9}$) depth profiling experiments were performed on these Bi-doping samples as well as the $Bi_2Te_3$(7 QLs)/FeTe heterostructure sample. In these Tof-SIMS experiments, Cs is used as the primary ion and BiCs and FeCs are used as the trace signals of Bi and Fe, respectively. Supplementary Fig. 4 shows the Tof-SIMS results. As shown in (a), it was found that both Bi and Fe signals across the whole FeTe layer just appear within a noise background, thus can be regarded as undetectable. In (b), (c), and (d), from the detected BiCs signals with a decreasing trend away from the surface, one can see that Bi is indeed successfully incorporated into the FeTe layer although quantitative concentrations cannot be determined due to the matrix effects and non-availability of a standard FeTe:Bi sample. It was expected that if any of these samples was able to turn FeTe into a superconductor, their transport properties should exhibit superconductivity at low temperature. However, their $R$ $vs$ $T$ behaviors measured down to the lowest measuring temperature as shown in Supplementary Fig. 3 do not show any superconductivity feature. From these results, one can conclude that our observed 2D superconductivity in the $Bi_2Te_3$/FeTe heterostructure is unlikely due to Bi doping.

**Possible intrinsic Te doping of FeTe.**

To investigate whether Te doping is the origin of 2D superconductvity, we have performed high-resolution EDS line-scan analysis on the $Bi_2Te_3$/FeTe heterostructure with the resulting scan shown in Supplementary Fig. 5, in which one can see that there is no detectable Te-rich region in the FeTe layer near the heterostructure interface since the relative amounts of Fe and Te both near and away from this interface remain the same. Moreover, a number of existing references[1-5] seem to conclude that FeTe layers fabricated by different techniques, in consequence with different stoichiometries, do not exhibit superconductivity. Thus, it is unlikely to expect our observed 2D superconductivity comes from intrinsic Te doping.

**Supplementary Table**

**Supplementary Table 1**

Dopant source temperatures used for growing the FeTe:Bi multilayers

|  | Bi source temperature of **FeTe:Bi # A** (°C) | Bi source temperature of **FeTe:Bi # B** (°C) | $Bi_2Te_3$ source temperature of **FeTe:Bi # C** (°C) |
|---|---|---|---|
| *(1)FeTe:Bi* | 265 | 322 | 370 |
| *(2)FeTe:Bi* | 290 | 345 | 380 |
| *(3)FeTe:Bi* | 300 | 360 | 390 |
| *(4)FeTe:Bi* | 310 | 390 | 400 |
| *(5)FeTe:Bi* | / | 405 | 410 |

**Supplementary Figures**

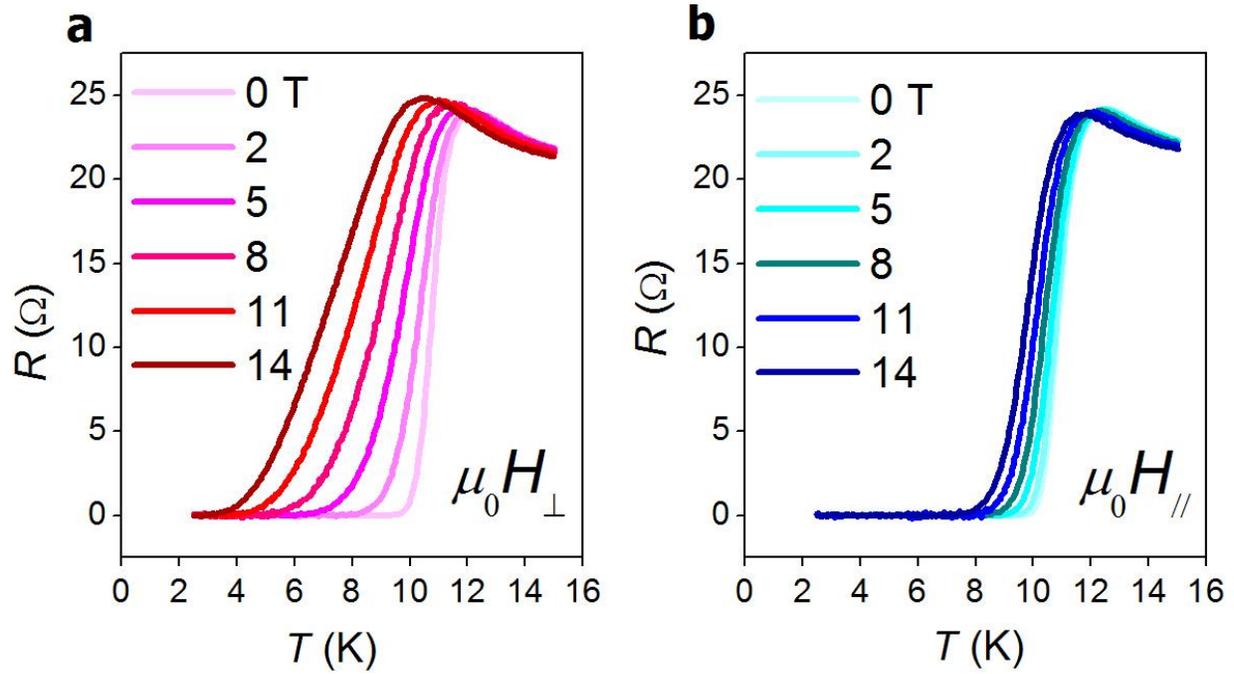

**Supplementary Figure 1**

$R(T)$ characteristics of Sample A [$Bi_2Te_3$(7 QLs)/FeTe heterostructure] measured under different magnetic fields in directions of (a) perpendicular and (b) parallel to the interface.

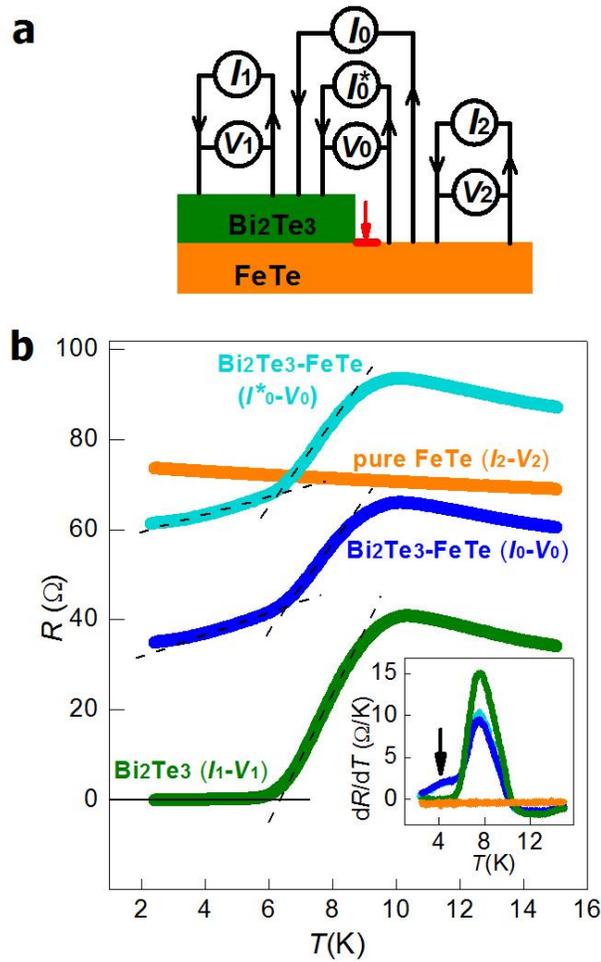

**Supplementary Figure 2**

Transport measurements of Sample B with 7 QLs of $Bi_2Te_3$ on half of the FeTe surface. (a) Schematic diagram of the sample structure and measuring methods. (b) Temperature-dependent resistances $R(T)$ obtained from different regions of this sample. The inset shows the corresponding $dR/dT$ curves.

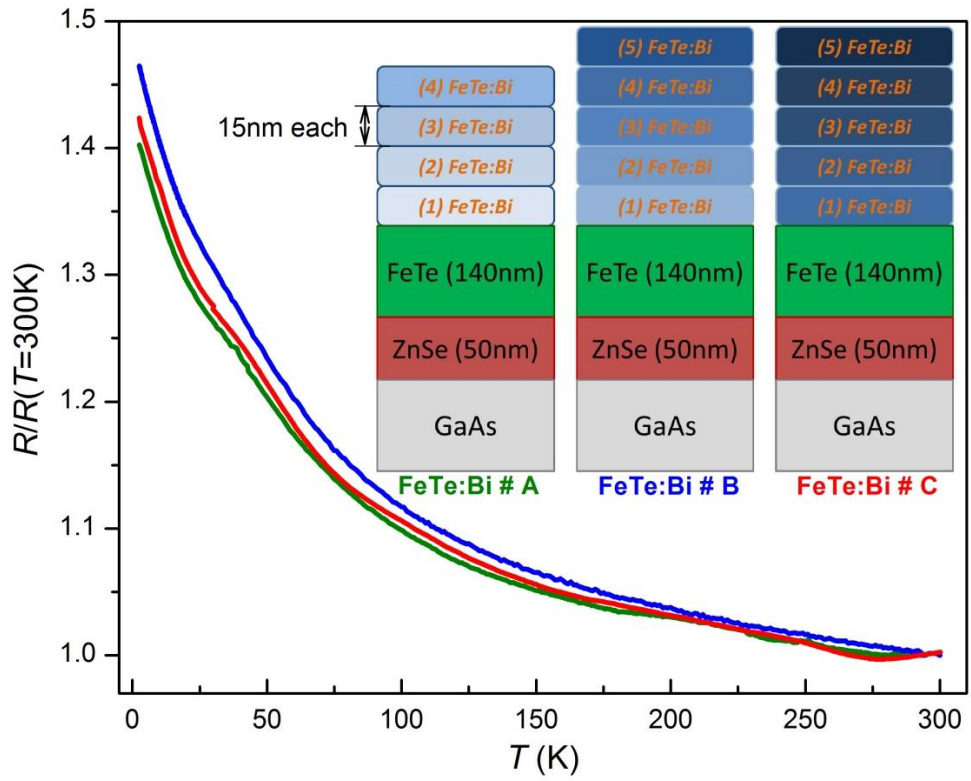

**Supplementary Figure 3**

$R(T)$ characteristics of samples FeTe:Bi#A, #B and #C. The inset shows their schematic sample structures.

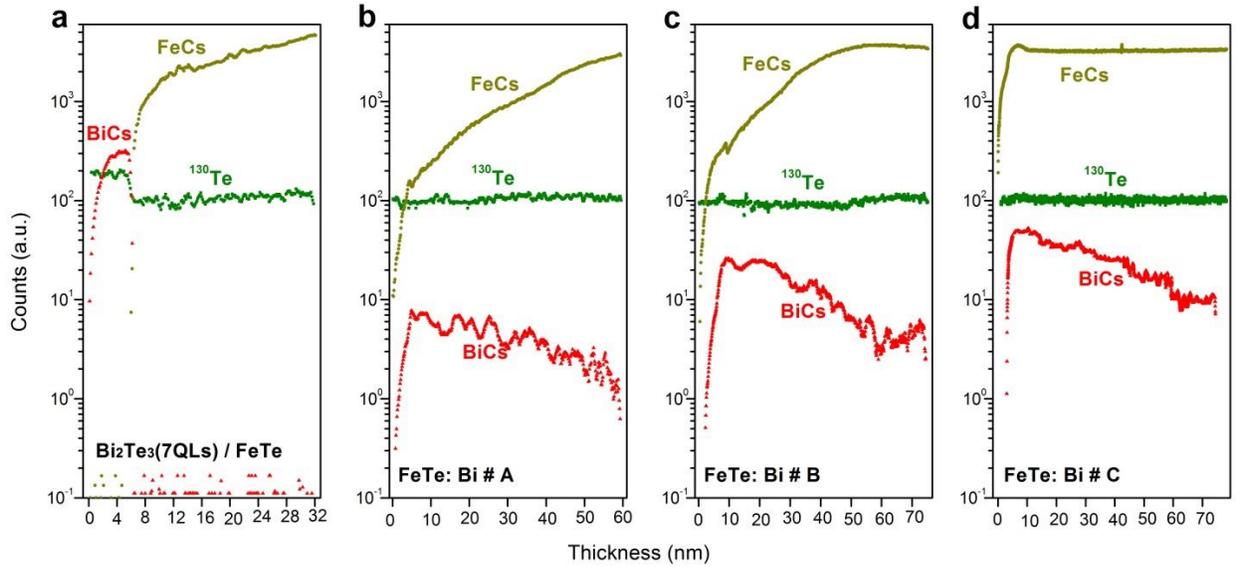

**Supplementary Figure 4**

Tof-SIMS depth profiles of (a) a $Bi_2Te_3$(7 QLs)/FeTe heterostructure, (b) FeTe:Bi #A, (c) FeTe:Bi #B, and (d) FeTe:Bi #C. The variation of FeCs signals and BiCs signals at the surface of the samples are due to surface oxidation. Sample FeTe:Bi #C suffers this variation much less as it was grown recently, other samples have been stored in air for more than a few months before performing the Tof-SIMS measurements.

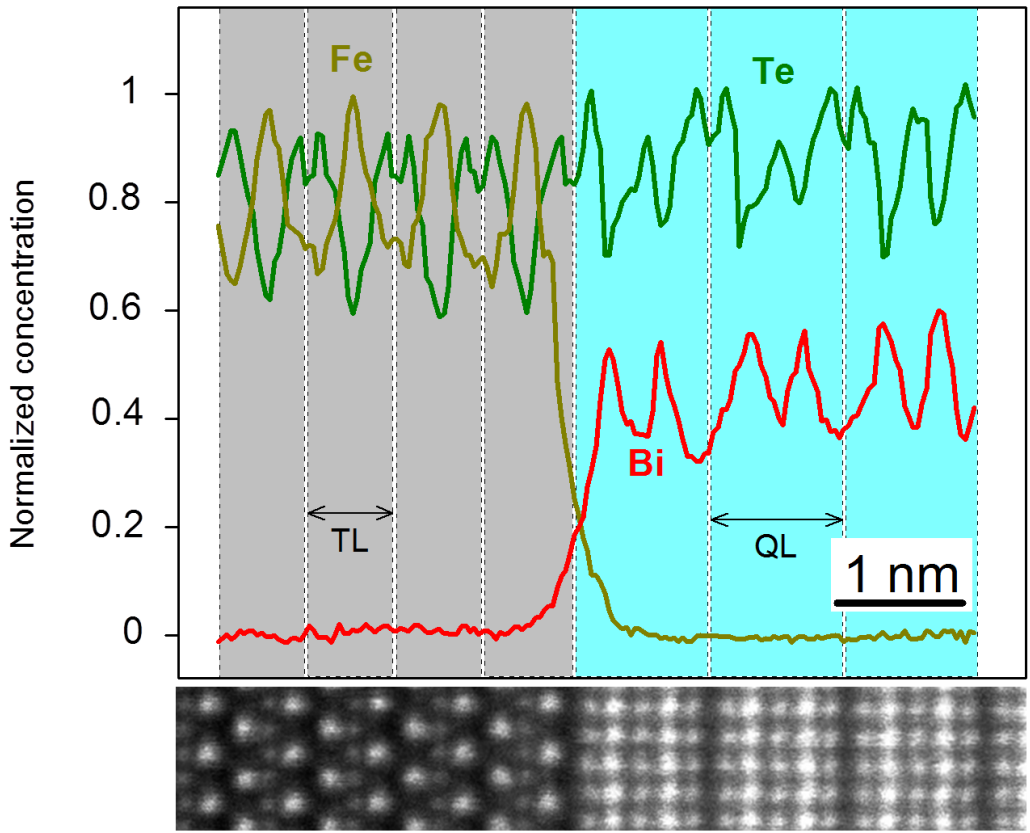

**Supplementary Figure 5**

Normalized concentrations of Bi, Te and Fe obtained by high-resolution EDS line-scan near the $Bi_2Te_3$/FeTe interface.

**Supplementary References**

1    Aswathy, P., Anooja, J., Sarun, P. & Syamaprasad, U. An overview on iron based superconductors. *Supercond. Sci. Technol.* **23**, 073001 (2010).

2    Mizuguchi, Y. & Takano, Y. Review of Fe Chalcogenides as the Simplest Fe-Based Superconductor. *J. Phys. Soc. Jpn.* **79**, 102001 (2010).

3   Chen, G. F. *et al.* Electronic properties of single-crystalline $Fe_{1.05}Te$ and $Fe_{1.03}Se_{0.30}Te_{0.70}$. *Phys. Rev. B* **79**, 140509(R) (2009).

4   Mizuguchi, Y., Tomioka, F., Tsuda, S., Yamaguchi, T. & Takano, Y. FeTe as a candidate material for new iron-based superconductor. *Physica C: Superconductivity* **469**, 1027-1029 (2009).

5   Zhang, C. *et al.* Pressure-induced lattice collapse in the tetragonal phase of single-crystalline $Fe_{1.05}Te$. *Phys. Rev. B* **80**, 144519 (2009).